\documentstyle [prd,aps,twocolumn]{revtex}
\newcommand{\la}[1]{\label{#1}}
\newcommand{\n}{{\bf n}}
\newcommand{\vecx}{{\vec{\bf x}}}

\newcommand{\be}{\begin{equation}}
\newcommand{\ee}{\end{equation}}
\newcommand{\ba}{\begin{eqnarray}}
\newcommand{\ea}{\end{eqnarray}}
\newcommand{\bastar}{\begin{eqnarray*}}
\newcommand{\eastar}{\end{eqnarray*}}
\newcommand{\half}{{1 \over 2}}
\newskip\humongous \humongous=0pt plus 1000pt minus 1000pt

\newif\ifdtup

\relax

\begin{document}
\title  {(META)STABLE CLOSED VORTICES IN 3+1 DIMENSIONAL
GAUGE THEORIES WITH AN EXTENDED HIGGS SECTOR
}
\bigskip

\author{ Antti J. Niemi$^{* \sharp}$, Kaupo Palo$^{\#}$ and 
Sami Virtanen$^{\dagger}$ } 

\address{
$^{*}$Department of Theoretical Physics, Uppsala University, 
P.O. Box 803, S-75108, Uppsala, Sweden
\\
$^{\sharp}$Helsinki Institute of Physics, 
P.O. Box 9, FIN-00014 University of Helsinki, Finland \\
$^{\#}$Evotec BioSystems AG, Instituudi tee 11, Harku EE76902, Estonia \\
$^{\dagger}$Material Physics Laboratory, Helsinki University of Technology,
Box 2200, FIN-02015 HUT, Finland \\
{\scriptsize \bf NIEMI@TEORFYS.UU.SE, PALO@EBI.EE, SAMI@FOCUS.HUT.FI } 
 }
\maketitle

\begin{abstract}
In gauge theories with an extended Higgs 
sector the classical equations 
of motion can have solutions that describe stable, 
closed finite energy vortices. Such vortices 
separate two disjoint Higgs vacua, with one of 
the vacua embedded in the other in a manner 
that forms a topologically nontrivial knot. The 
knottedness stabilizes the vortex against shrinkage
in 3+1 dimensional space-time. But in a world 
with extra large dimensions we expect 
the configuration to decay by unknotting. As an example we 
consider the semilocal $\theta_W \to \frac{\pi}{2}$ 
limit of the Weinberg-Salam model. 
We present numerical evidence for the existence of 
a stable closed vortex, twisted into a toroidal 
configuration around a circular Higgs vacuum at its core.
  
\end{abstract}
\vskip 0.2cm

\narrowtext
\bigskip
\noindent
{\bf 1. Introduction}
\vskip 0.2cm
In unified field theory models the existence of several 
degenerate Higgs vacua is generic \cite{higgs}. This 
degeneracy has numerous consequences, and leads to
the appearance of domain walls \cite{vilenkin} and 
vortices \cite{vach}. The role and properties of 
such configurations have been widely discussed.
They are expected to be highly relevant 
in a variety of problems in high energy 
physics and Early Universe cosmology.

The energy of a line vortex scales with its length and
a stable vortex is expected to have an infinite energy.
Any isolated, finite length vortex such as a vortex
loop in the standard abelian Higgs model with a single 
complex scalar field is expected to rapidly 
collapse. But here we propose that (meta)stable, or at 
least very long-lived finite length closed vortices 
can actually be present in unified field theory models with an
{\it extended} Higgs sector. Even though the energy of 
these vortices does scale in proportion to their length, 
they can still be prevented from shrinking by 
their topological non-triviality. This becomes possible when 
the vortex is deformed so that it forms a knot,  
separating two Higgs vacua which are tangled around 
each other in a nontrivial manner. The topological 
nontriviality of a knotted structure then 
provides a repulsive force which stabilizes the 
vortex and prevents it from collapsing.

An actual dynamical stability of a knotted 
finite energy closed vortex depends on the dynamical details 
of the underlying field theory model,
and must be verified separately by inspecting
the classical equations of motion. 
But we shall argue that with a properly extended 
Higgs sector dynamical stability can occur. 
The vortex becomes classically protected from a
collapse by a finite energy barrier, very much 
in analogy with the mechanism that has been 
introduced in the case of one dimensional 
models in \cite{toma}. Indeed, following \cite{toma} 
we propose that such a finite energy barrier 
persists even at low coupling, where semiclassical 
methods become reliable. This means that quantum 
mechanically the vortex can have a very long life-time, 
even though it may eventually decay by a tunneling 
process \cite{toma}. 

The sample field theory model that we consider 
in the following is related to the Weinberg-Salam 
model. It emerges when we take the limit where 
Weinberg angle $\theta_W \to \frac{\pi}{2}$. This
suggests that our configurations may have 
relevance to Standard Model physics, at 
energy scales which may become reachable in the  
future accelerators such as LHC. A detailed 
inspection of knotted vortices in 
realistic (supersymmetric) extensions of the 
Standard Model could then become rewarding.
For example, since a nontrivial knot is 
topologically stable exactly in three space 
dimensions, the existence of stable knotted 
vortices could provide a method for testing
the dimensionality of space-time at the TeV 
scale, an issue which is presently under 
an active debate \cite{anto},
\cite{lisa}. If extra dimensions 
are indeed visible to the Standard Model at TeV scales, 
it should be difficult to preserve the 
stability of TeV (or higher) scale knotted 
configuration. These configurations should 
rapidly decay, by unknotting themselves in 
the extra dimensions. Instead of knotted
vortices of the type discussed here, in 
an extension of the Standard Model 
with TeV scale extra dimensions it should then
become relevant to inspect the stability of knotted 
configurations that are formed by entanglements of
submanifolds with co-dimensionality 
two in the extended world.

\vskip 0.2cm
\noindent
{\bf 2. The Model}
\vskip 0.1cm
The topological nontriviality of a knot is characterized
byt its self-linkage, and the ensuing linking 
number coincides with the Hopf invariant \cite{ati}. 
This is an integral invariant that relates self-linking to 
the $\pi_3(S^3) \sim \pi_3(S^2)$ homotopy classes.
In the case of static and localized configurations such as a 
knot, we may compactify $R^3 \to S^3$. This suggests that
in the case of a Yang-Mills -Higgs model, a knotted 
vortex might be present when the Higgs vacuum sector 
allows for a nontrivial $S^3 \to S^3$ or $S^3 
\to S^2$ mapping. This is possible when
the symmetry group of the Higgs vacuum sector 
contains a $SU(2)$ subgroup. A familiar example is the
Weinberg-Salam model, and we shall exemplify our ideas
by considering its simplified version, the so-called semilocal
{\it i.e.} $\theta_W = \frac{\pi}{2}$ limit of the Weinberg-Salam
model \cite{vach}. This is an abelian gauge theory with a 
two component complex Higgs scalar and renormalizable Lagrangian
\be
L \ = \ - \frac{1}{4} 
F_{\mu\nu}^2 \ + \ \bar D_\mu Z^\dagger D_\mu Z
- \lambda ( Z^\dagger Z - 1 )^2
\la{model1}
\ee
where $D_\mu = \partial_\mu - i A_\mu$ 
is the U(1) covariant
derivative and
\be
Z \ = \ \left( \matrix{ \varphi_1 \cr
\varphi_2} \right) \ = \ |\Phi| \cdot
\left( \matrix{ \phi_1 \cr \phi_2 \cr
\phi_3 \cr \phi_4 } \right)
\la{Z}
\ee
the Higgs field, and we normalize $\sum \phi_i^2  =  1 $. 
The Higgs vacuum manifold of (\ref{model1}) is then 
characterized by $|\Phi| = 1$ which determines the topology 
of $S^3$. In this model Higgs mechanism occurs, 
the gauge field $A_\mu$ and one 
component of the complex $\varphi_a$ combine into a massive 
vector field. In addition we have a massive
scalar and two Goldstone bosons.   

The model (\ref{model1}) is known to support straight, 
infinite-length, infinite-energy line vortices 
as stable classical solutions to its equations of 
motion, and the properties of these vortices have been 
widely discussed \cite{vach}. In the following we argue that 
the model (\ref{model1}) also supports static, finite length
and energy knotted vortices as stable solutions. The
vortex now appears as a domain wall that separates two 
different Higgs vacua. Modulo global $SU(2)$
rotations we can select these two Higgs vacua so 
that in $R^3$ they coincide with the preimages of 
the two $S^1$, which are determined by setting 
either $|\varphi_2|(\vecx) = 1$ or $|\varphi_1|(\vecx)  
= 1$ (with $|\Phi|(\vecx) = 1$). In the sequel
we shall denote these two vacua as $S^1_{out}$ and 
$S^1_{in}$, respectively. For a potentially 
stable closed vortex to occur, these two vacua should be linked
with a nontrivial linking number which is computed 
by the Hopf invariant $Q_H$ \cite{nature} 
\be
Q_H \ = \ \frac{1}{12\pi^2} \int \epsilon_{ijkl} \
\phi_i d \phi_j \wedge d \phi_k \wedge d \phi_l
\la{hopf}
\ee
We note that this coincides 
with the $\pi_3(S^3)$ winding number of
the field $\phi_a$.

We argue that for a nontrivial $Q_H$ the vortex  
which separates the two Higgs vacua can be classically 
(meta)stable, at least for some range of values for 
the coupling constant $\lambda$. 
Indeed, the classical equation of motion 
obtained by varying (\ref{model1}) {\it w.r.t.}
$A_i$ is (we consider static configurations 
in the $A_0=0$ gauge)
\be
A_i \ = \ - \half \frac{1}{Z^\dagger Z} (
\partial_i Z^\dagger Z - Z^\dagger \partial_i Z )
\ + \ \frac{1}{4} \frac{1}{Z^\dagger Z} \  \partial_j
F_{ij}
\la{Aeq}
\ee
By sending $\lambda \to \infty$ we force $Z^\dagger Z 
= |\Phi| \equiv 1$, and if we define a three component 
unit vector  by $ \hat \n  =  Z^\dagger {\hat \sigma} 
Z $ and use (\ref{Aeq}) to eliminate $A_i$ 
from (\ref{model1}), we find  that 
(\ref{model1}) reduces to
\be
L \ \to \ |\partial_\mu \hat \n |^2 \ + \ 
\frac{1}{4e}
( \hat \n \cdot d \hat \n \times d \hat \n )^2
\ + \ ({\rm higher \ derivatives})
\la{fad}
\ee
The first two terms that we have presented 
in (\ref{fad}) reproduce the model studied 
in \cite{nature}, \cite{dur}. There, it has been 
established that (\ref{fad}) admits stable 
knotlike solitons with a nontrivial Hopf invariant.
Those results then suggest, that at 
least for large 
values of $\lambda$ \cite{remark1} 
the model (\ref{model1}) should either support 
stable knotted configurations with nontrivial $Q_H$, or 
then it should describe 
metastable knotted configurations with a 
life-time that approaches infinity as $\lambda \to 
\infty$. 

\vskip 0.2cm
\noindent
{\bf 3. Stable Toroidal Vortices }
\vskip 0.1cm
The previous arguments are suggestive but not  
sufficient to conclude that for finite, even for weak
$\lambda$ the model (\ref{model1}) could support stable 
closed vortices. The equations of motion are highly  
nonlinear and for finite coupling
the higher order terms in (\ref{fad})
can not be ignored. Consequently an explicit construction 
with a finite $\lambda$ becomes imperative.
For this we remind that (\ref{model1}) supports 
stable, infinitely long line 
vortices \cite{vach}, \cite{remark1}. 
If we construct a finite energy vortex in (\ref{model1})
by cutting a piece of the line vortex studied in
\cite{vach} which we then
deform into the shape of a finite radius torus by joining
the ends, the resulting 
configuration becomes unstable against shrinkage.
However, if we form such a toroidal  
vortex ring by first twisting the line
vortex once around its core before joining its ends, 
the nontriviality of the twist might produce a 
repulsive interaction that stabilizes the configuration 
against shrinkage \cite{nature}. In order to form an 
appropriate configuration that allows for the introduction
of a nontrivial twist, we recall that the present model 
reduces to the standard abelian Higgs model in the limit 
where we truncate one of the two complex fields $\varphi_a$. 
In the standard abelian Higgs model 
the number of degrees of 
freedom is insufficient for describing 
a nontrivial twist around the core
of a vortex, and a toroidal configuration is unstable. 
But in the present case the Higgs sector
is extended with the broken phase containing two additional
Goldstone bosons. This ensures that 
the number of degrees of freedom is now sufficient for 
describing a nontrivial twist around the core of
a vortex. For this, we form a toroidal configuration 
in such a manner that outside and inside of a toroidal surface
we have a different asymptotic large distance Higgs vacuum 
of the standard line vortex in the conventional abelian Higgs model. 
This means that outside of the toroidal surface 
we select {\it e.g.} the Higgs vacuum $S^1_{out}$
which is characterized by $|\Phi| = |\varphi_2| = 1$,
and inside of the torus we select the Higgs vacuum 
$S^1_{in}$ with $|\Phi|=|\varphi_1|=1$. These two 
Higgs vacua become then separated by a toroidal domain 
wall configuration that interpolates 
between the two vacua. We identify this domain wall
as our closed vortex, wrapped around 
the toroidal surface.  

The two Higgs vacua $S^1_{in}$ and $S^1_{out}$ are now linked
in a nontrivial manner. If we assume that $|\Phi| \not= 0$ 
everywhere, the Hopf invariant (\ref{hopf}) is everywhere 
well-defined and it computes the linking number of these two 
Higgs vacua. A non-vanishing Hopf invariant provides 
topological stability for the toroidal vortex that interpolates
these Higgs vacua, the vortex can unwind 
only if the norm $ |\Phi|$ in (\ref{Z}) vanishes 
for some $\vecx \in R^3$ so 
that the Hopf invariant becomes ill-defined. 
The Higgs potential provides an energy 
barrier that prevents $|\Phi|$ 
from vanishing. For finite $\lambda$ 
this energy barrier is finite.
It increases with an increasing $\lambda$ 
and becomes infinitely high in
the limit where $\lambda \to \infty $. For 
finite values of $\lambda$ 
the vortex can then be stable, provided the 
equations of motion indeed
restrict the Higgs field so that it is 
everywhere constrained to $|\Phi(\vecx)| > 0$. 
In this manner classical (meta)stability 
becomes a dynamical issue, it can be resolved 
only by actually solving 
the equations of motion.

Following \cite{toma} we note that 
since the energy barrier is finite,
quantum mechanically the vortex 
may still decay. It may unwind itself
by barrier penetration, when 
the Higgs fields fluctuates
so that for some region of 
space $|\Phi(\vecx)|$ vanishes.
However, if the classical solution 
has $|\Phi(\vecx)| \approx 1$ 
as we expect {\it e.g.} for 
large $\lambda$, the vortex  
has an exponentially long quantum 
mechanical lifetime. Indeed, since the
present situation is quite 
similar to the one dimensional case
studied in \cite{toma}, we 
expect that also in the present case
a classical solution remains 
stable even for relatively small 
values of $\lambda$.

\vskip 0.2cm
\noindent
{\bf 4. A Numerical Solution}
\vskip 0.1cm
The equations of motion for (\ref{model1}) are highly
nonlinear, to the extent that an analytical investigation
of an actual vortex configuration appears entirely hopeless. 
Consequently we resort to numerical methods.
We have performed an extensive search for a (locally stable) 
vortex configuration in a toroidal shape. 
For this, we specify a torus on the $(x,y)$
plane, oriented so that the toroidal symmetry 
axis coincides with the $z$ axis. With $(\rho,z,\psi)$
cylindrical coordinates, we use rotation invariance
to argue that the most general $SO(2) \times SO(2) $ 
invariant Ansatz for the Higgs field can be represented
in the functional form
\be
Z \ = \ \Phi(\rho,z) \cdot \left( \matrix{ \cos k\psi 
\cdot  \sin \half\theta(\rho,z) \cr \\
\sin k\psi \cdot \sin \half \theta (\rho,z) \cr \\
\cos \vartheta (\rho,z) \cdot \cos \half \theta (\rho,z) \cr \\
\sin \vartheta (\rho,z) \cdot \cos \half \theta (\rho,z) }
\right)
\la{Ansatz}
\ee
where $0 \leq \vartheta < 2 \pi n$ and $0 \leq \theta < \pi$
and $k, \ n$ are integers.
For the gauge field $A_i$ the rotation invariance 
restricts the components into functions of $(\rho,z)$ only. 
With this Ansatz the static equations of motion for (\ref{model1}) 
reduce into a set of coupled equations for six unknown functions, 
which we then study numerically. We use the following boundary 
conditions: Both at $\rho = 0 $ and for $\rho$ large
$\theta(\rho)$ vanishes and $\Phi(\rho) = 1$, corresponding
to our Higgs vacuum configuration $S^1_{out}$. For some point
at a distance $\rho = \rho_c$ we 
have $\theta(\rho_c , 0) = \pi$ with $\Phi(\rho_c , 0) = 1$,
corresponding to our Higgs vacuum configuration $S^1_{in}$.
These boundary conditions determine a toroidal
vortex wrapped around the circle $S^1_{in}$, it appears as
a domain wall that separates the Higgs vacuum $S^1_{in}$ 
from the asymptotic Higgs vacuum $S^1_{out}$.
The Hopf invariant of this configuration (when $\Phi \not= 
0$ everywhere) is $Q_H = k \cdot n$ so that the vortex is
topologically stable: It wraps the torus $k$ times, and
twists $n$ times around its center. We expect that 
this twist produces a repulsive force that balances 
the vortex against shrinkage, resulting 
into a (meta)stable configuration.

We have employed our Ansatz to perform very extensive numerical 
simulations to search for a toroidal vortex configuration. 
In our simulations we have used the techniques described 
in \cite{nature},  by extending the Hamiltonian equation 
into a parabolic flow equation with respect
to an auxiliary time variable and then following the 
flow towards a fixed point of the Hamiltonian.
We have found definite convergence towards 
torus-shaped configurations, that appear stable under 
our parabolic flow. This indicates that the 
Hessian is positive definite when evaluated 
at the fixed point, which is
a manifestation that our final configuration 
is indeed classically stable. As an example, 
in the figure we describe a solution for $k=2$ and $n=1$. 
In this figure we plot the profile for the angular
variable
\be
\Theta(\rho,z) \ = \ 2\arctan \left( \frac{|\varphi_1|}{|
\varphi_2|}\right)
\la{thet}
\ee
It varies between $\Theta_{in} = \pi$ at $\rho_c \approx
3.8$ and $\Theta_{out} = 0$ outside of the configuration,
so that we indeed have a domain wall configuration 
that separates the two Higgs vacua $S^1_{in}$ 
and $S^1_{out}$ in the expected manner. In this figure
we have also plotted the contours for the energy density,
which we find peaks near $\rho_c \approx 3.3$
for this particular configuration. We note that
when we approach the $z$-axis, the energy density 
vanishes. The qualitative
behaviour of the energy density is consistent with the
expected behaviour of the Higgs field near the
two vacua $S^1_{in}$ and $S^1_{out}$. When we approach
the Higgs vacuum $S^{1}_{in}$, we expect that
$|\Phi| \approx 1 + {\cal O}( |\rho - \rho_c|^n )$ so
that with $n=1$ the 
derivative terms in (\ref{model1}) yield a
finite contribution to the energy density even at
$\rho = \rho_c$, and when we approach 
the $z$-axis we expect $|\Phi| \approx 
1 + {\cal O}( \rho^k )$ so that with $k=2$ the energy density 
vanishes on the $z$-axis.
 
The equations of motion, even with the simplified Ansatz
(\ref{Ansatz}), are highly complex and practical 
simulations become very time consuming. A typical run
consumes over a hundred hours of CPU time, when we use 
a finite element method implemented with the PDE2D 
program \cite{sewell} and a single EV56 processor 
in a Digital AlphaServer 8400. 
We have inspected the solutions to the classical equations of 
motion for various values of the coupling $\lambda$, 
and found evidence of convergence even for moderately
small values of $\lambda \sim {\cal O}(1)$. However, for
such small values of $\lambda$ the simulations become increasingly
involved, and numerical convergence becomes increasingly sensitive to 
the choice of an initial condition.  This is expected:
When we use the analogy with \cite{toma}, we conclude 
that for convergence towards a stable vortex
we need to locate 
an initial configuration which can not "slip off the tip
of the Mexican hat", using the analogy with the
discussion in \cite{toma}. When $\lambda$ 
decreases the choice of a suitable initial configuration 
becomes then increasingly more difficult, and the ensuing simulation
similarly increasingly more time consuming. Unfortunately, at the 
moment we do not have access to sufficiently effective computers
and numerical methods to locate a lower bound $\lambda_c$
for the existence of (locally) stable solutions in a reliable 
manner.  Indeed, it appears that a detailed numerical 
investigation of the properties of our vortices must 
be postponed to the future \cite{lk}, when more powerful computers 
become available. However, we expect that these simulations
will eventually reveal a qualitative behaviour which is very
similar to that found in the lower dimensional analog \cite{toma}.
Among the conclusions in \cite{toma} are, that the 
configurations should be generic in gauge models with an extended 
Higgs sector. Moreover, the solutions can be stable even in the 
weak coupling limit where semiclassical methods become reliable. 
If these properties indeed persists in 3+1 dimensions,
we can expect that our vortices are generic in models with 
an extended Higgs sector, even in the low coupling limit where 
semiclassical methods become reliable. If so, they should have 
relevance in a number of physical scenarios, including 
applications to LHC physics and Early Universe Cosmology.

\vskip 0.2cm
\noindent
{\bf 5. Conclusions}
\vskip 0.1cm
In conclusion, we have argued that closed finite energy 
knotted vortices 
may be generic in gauge theories with an extended Higgs sector.
These vortices appear as domain walls that separate two 
different Higgs vacua. They are topologically nontrivial if  
the vacua are tangled into knots with a nontrivial Hopf invariant. 
Actual dynamical stability of a closed vortex depends on the details of 
the model. For this the Higgs potential should provide a sufficiently 
high, "tip of the Mexican-hat" type energy barrier 
so that in combination with the derivative terms 
it prevents the vortex from unwinding. Studies of an analogous
behavior in one dimension suggests that closed vortices could be generic
in theories with an extended Higgs sector, and their stability 
could persist even at low coupling where semiclassical
methods become reliable. If present, these configurations
should have relevance to LHC physics and Early Universe Cosmology.
In particular, since knots are topologically stable only when 
embedded in three space dimensions, the very existence 
of (meta)stable knotted vortices could develop into a viable 
method for experimentally testing the dimensionality of the 
World, as seen by the Standard Model at TeV scales.

\vskip 0.2cm
\noindent
{\bf Acknowledgements} A.N. thanks Ludvig Faddeev for many
discussions. A.N. also thanks T. Tomaras for drawing attention
to the results in \cite{toma} and for suggesting that
there might be a similarity, and for discussions on the
potential relevance of knotted vortices to Standard Model
Physics. We are grateful to the Center for Scientific 
Computing in Espoo, Finland for providing us with an 
access to their computers. The research by A.N. was partially supported 
by NFR Grant F-AA/FU 06821-308.

\vfill\eject
\begin{flushleft}
{\bf Figure Caption}
\end{flushleft}
\vskip 0.0cm
\noindent
{\bf Figure:} A typical profile (\ref{thet}) for a stable 
configuration with $n=1$ and $k=2$ and $\lambda = {\cal O}(10)$.
At the center of the torus, here at a distance $R \approx 3.8$ from
the $z$-axis,  we have $\Theta = \pi$ corresponding
to the Higgs vacuum $|\varphi_1|=1$ and $\varphi_2 = 0$. 
Outside of the torus we approach the Higgs vacuum with
$|\varphi_2|=1$ and $\varphi_1 = 0$. The closed vortex
interpolates between these two Higgs vacua. Also in the 
figure we have the contours for energy density, which 
peaks near $\Theta = \pi$ but 
at a somewhat smaller distance $R$ from the $z$-axis.


\begin{thebibliography}{99}
\bibitem{remark1} Note that unlike in the present case and the
case studied in \cite{toma}, the stability of 
the semilocal string \cite{vach} occurs for small $\lambda$.
\end{thebibliography}
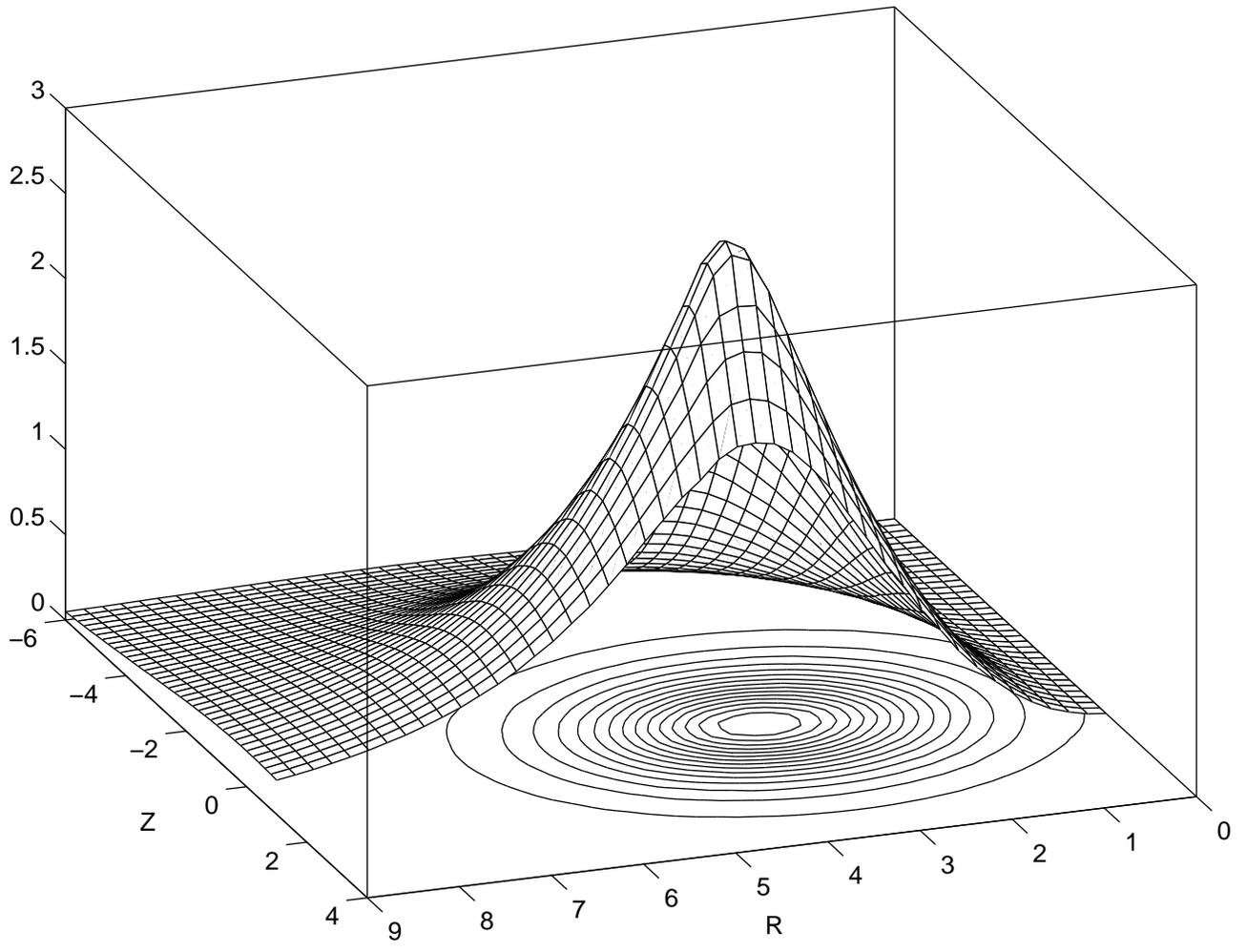
\end{document}